KJSET                                                                                                              Research
Article

# On-site Energy Utilization Evaluation of Telecommunication Base Station: A Case Study of Western Uganda

**Aceronga Kwocan[1], Mohammed Dahiru Buhari [1,2], Kelechi Ukagwu[1], and Jonathan Serugunda[3]**

*Department of Electrical, Telecommunication and Computer Engineering, Kampala International University, Uganda[1]*
*Electrical and Electronic Engineering Department, Abubakar Tafawa Balewa University, Bauchi, Nigeria[2]*
*Electrical Engineering Department, Makerere University, Uganda (MUK)[3]*
*kwocan.aceronga@studwc.kiu.ac.ug[1], ukagwu.john@kiu.ac.ug[1], dbmohammed@atbu.edu.ng[2],*
*serugthan@gmail.com[3]*

*Corresponding Author: kwocan.aceronga@studwc.kiu.ac.ug[1]*

## Abstract

*Due to the widespread installation of Base Stations, the power consumption of cellular communication is increasing rapidly (BSs). Power consumption rises as traffic does, however this scenario varies from geolocation to geolocation because sites in rural and urban areas have variable traffic loads. Therefore, in order to address various power consumption issues, it is necessary to analyze these sites and offer validate data that network operators can employ. This study took into account the impact of traffic load on the energy consumption both in rural and urban locations in western Uganda because prior models did not adequately account for the impact of traffic load on both rural and urban sites. Regression models are used to examine these effects of traffic load on power consumption. Based on measurements taken for twenty-eight days in a row at six urban and rural areas, linear models have been presented. The findings showed that both rural and urban BTS were well-fitted by the suggested linear models. Depending on the layouts of the sites, it was found that energy consumption varied along with traffic, with the number of transceivers present having an impact on both the traffic load and energy consumption.*

### Nomenclature and units

| | |
|---|---|
| 2G | Second generation of network |
| 4G | Third Generation of Network |
| TRX | Transceiver |
| BS | Base Station |
| UMTS | Universal Mobile Telecommunication System |

## 1.0 Introduction

For telecom firms around the world, including in underdeveloped nations like Uganda, high energy consumption in base stations (BTS) of telecommunication has long been an issue (Lubritto et al; 2008). This significant energy usage keeps rising daily and makes it difficult for network providers to control their market. Due to the increased network traffic in urban areas and in areas without adequate grid coverage, where network operators are required to install their own power sources (Lubritto et al; 2008) setting up their own generators for the production of the energy to the equipment at site, the growth of this energy consumption is enormous.

Because of this, attention must be given and energy consumption in the communications base station must be stabilized in order to solve the energy consumption issue. A telecom network is similar to an eco-system in that one cannot simply implement any energy-saving measures without considering the effects on the other system components, as (Roy noted in 2008). In order to create a stabilized model for on-site energy usage, a thorough survey and investigations of the various models were conducted.

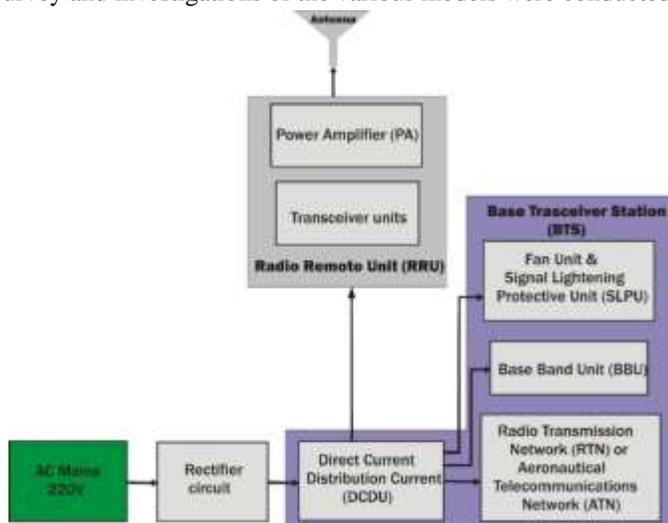

**Figure 1** Block diagram of the site.

Since the sites we visited were all outdoors, there wasn't much more equipment consuming the energy besides the radio units and the base band units, therefore we constructed regression models to provide a better power consumption both at the base station (Obinna&Osawaru, 2020). Power Systems are one of the key components of telecommunications technology, allowing for significant financial resource savings for managing mobile communications systems as well as achieving "sustainable" development goals. (Lubritto et al; 2009). In other words, reducing energy usage is a practical opportunity that can help network operators and subscribers in Uganda and around the world. Fighting against global warming is especially important because communications networks produce 1% of the world's energy (Zhou et al; 2016). This is the same as the energy use of

15 million homes and the CO2 emissions of 29 million cars combined.

Due to the increased interest in the telecom industry, particularly in the western region where there are more grid coverage zones, more base stations are currently required in Uganda. Network operators must establish their own (Amankwah &Amoah, 2015). Power source for the generation of electricity for the sites to guarantee a higher quality of service (QoS) for subscribers in the coverage zone everywhere and at all times. energy use has increased as a result of network technology. The new Fifth Generation (5G) network system is one of the factors contributing to the high energy consumption in the mobile communication industry (Amankwah & Amoah, 2015). Despite the fact that the Fourth Generation of Network (4G) system uses a lot of power, and the Fifth Generation of Network (5G) system is designed to operate under a low consumption power, the 5G system is designed to use a spectrum that requires numerous base stations and node Bs and typically results in three times the power consumption (Amankwah &Amoah, 2015). However, as this study included traffic calls rather than data, 5G and 4G were not our primary concerns. With an emphasis on western Uganda, the current study examined the on-site energy consumption in base stations of telecommunication for Airtel locations in Uganda.

## 2.0 Materials and Methods
### 2.1 Materials

In this work, the following materials were used to collect data: Clamp meter and Multimeter and a laptop to save these data**.**

### 2.1.1 Clamp Meter

The DC current flowing through the cables connecting the Radio Units, which are interfaced to each sector on sites, was measured for these sites using a clamp meter. Hourly measurements were made at each cell location over the course of a month, or four (4) weeks, as illustrated in Fig. 2.

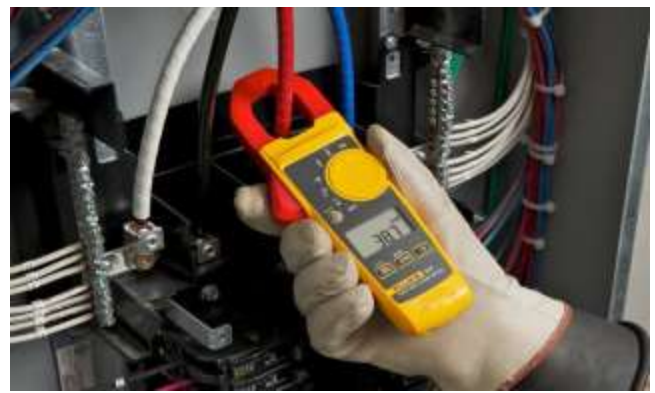

**Figure 2** Clamp meter.

### 2.1.2    Multimeter

In order to extract or to calculate the value of the utilized energy, voltage and current values need to be collected, therefore, the need to use a Multimeter to collect the voltage of each equipment at site. The voltage across each equipment needs to be taken for analysis. This measurement was done hourly for the period taken to collect data. Illustration is seen in Fig. 3.

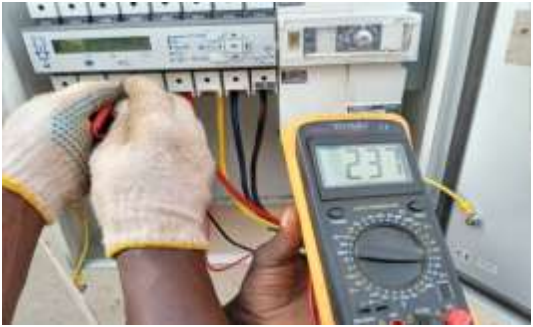

**Figure 3** Multimeter.

## 2.2    Methods

### 2.2.1    Data Collection

Data for this study was collected from base stations in the forementioned research locations. Data collection took place at 6 base stations in the Bushenyi, Ishaka region and Mbarara city, each of which had a unique site layout that allowed for adequate analysis and comparison. As soon as the plan was presented, this was done. At each base station location visited during this step, a clamp meter instrument was employed to permit the collection of samples. In this sample collection, current and voltage values were measured, and from them, power, current, and energy were determined.

This study selected a clamp meter and a multimeter to collect data values because, site being on air, could not be interrupted, more so the study needed the site to be on air, so as to perform analysis. However, some equipment such as the rectifier had a table displaying the values consumed over a period of time, this also helped in getting the values straight from the displayed screen rather than making measurements, so as to ease the task of data collection.

### 2.2.2    Calculation of power values from the collected Voltages and Current values

From the sample collected, that is, current and voltage values, Power value was determined (Dahal et al; 2017). These numbers enabled this study to conduct analysis on the on-site sample for the outdoor sites visited during the collection of data. Since we are only looking at DC circuitry in this study, from power equation (Obinna&Osawaru, 2020) for DC is given in equation (1).

$$P_{(W)} = V_{(V)} \times I_{(A)} \qquad (1)$$

The Average Voltage from 6am to 6pm is given by;

$$V_{(av)} = \frac{\sum_{t=0}^{12}(V)}{12} \qquad (2)$$

The Average Current from 6am to 6pm is given by;

$$I_{(av)} = \frac{\sum_{t=0}^{12}(I)}{12} \qquad (3)$$

Therefore, the Average Power consumption from 6am to 6pm is given by;

$$P_{(av)} = V_{(av)} \times I_{(av)} \qquad (4)$$

We considered the average power consumption in order to achieve our objectives. This came due to analysis that is based on the traffic loads of the sites thereafter, comparison was made between average of the power consumed at each base station and the traffic load so as to understand the behavior of energy consumption vis a vis traffic lad.

### 2.2.3    Analysis of data

Power consumption vs Traffic load Analysis has been made on the collected values, for each site and for each day in the 28 days of measurements. This was successfully done in MICROSOFT EXCEL. Out of these values nights weren't considered since we were not allowed access to site at night. Nevertheless, a less significant value of utilized energy is consumed at night hours, in between the hour of 10pm up to 5am. This information was given to us from the data control center of the Huawei/Airtel. This means that all the instantaneous average values for power and traffic load were scattered in Excel in order to determine if or if not, these values can fit for a linear regression model.

### 2.2.4    Data comparison with standard energy consumption from Airtel, ATC

A typical power consumption for each equipment at site has been provided by Airtel company, in order for us to use it and compare the data we have to see if it matches the standards required by this company. According to the analysis, we came to know that these site's energy consumption is in line with the standards. However, the instability of instantaneous power consumption needs to be looked at. This means that at some point, there is a higher energy consumption at site as compared to what is expected to come out of each site. So, the need for this study is to propose models for the sites. From the values gotten from the calculations of the power consumption and average power consumption, we know that the average energy at these base stations is linearly consumed. Therefore, the need to make a linear regression model for the sites.

#### 2.2.4 Data validation

A linear regression model was developed to validate data. Our data being linear, this regression gives us a clear view on how best power can be managed at the base station of telecommunication. For each site and each technology, a linear regression model has been developed as mentioned in the objectives of this study.

#### 2.2.4.1 Power consumption model Parameters

The table below provides a summary of the parts of our BS as well as the parameters governing their power consumption (James et al; 2016).

**Table 1** Power consumption Model parameters.

| Components | Power consumption parameter |
|---|---|
| Base Band Unit | $P_{BBU}$ |
| Radio Remote Unit | $P_{RRU}$ |
| Rectifier | $P_{REC}$ |
| Fan unit | $P_{FU}$ |
| Fluorescent Bulb | $P_{FB}$ |
| Base Station Power | $P_{BTS}$ |

Equation (2.5) was used to calculate the power used by each component at site from the measured currents and the measured voltages.

$$P_{(W)} = V_{(V)} \times I_{(A)} \qquad (5)$$

The total power used by a BTS is calculated as the sum of the power used by each of its parts and can be calculated as follows:

$$P_{BTS} = P_{BBU} + P_{RRU} + P_{REC} + \sum_{k}^{k} P_{FB} + \sum_{m}^{m} P_{FU} \qquad (6)$$

The power utilized at a base station PBTS was separated into two categories: traffic dependent and traffic independent since the measured current values for some base station components did not vary with traffic load (James et al; 2016). Traffic load does not affect the measured values of the following model components; fan unit, fluorescent bulbs, and BBU. However, the RRU values were influenced by traffic load. This is the reason we name it traffic independent component. Equation (6) illustrates the total base station power.

$$P_{BTS} = P_{Trafficindependent} + P_{Trafficdependent} \qquad (7)$$

Hence,

$$P_{Trafficindependent} = P_{BBU} + P_{REC} + \sum_{k=1}^{k} P_{FB} + \sum_{m=1}^{m} P_{FU} \qquad (8)$$

$$P_{Trafficdependent} = P_{RRU} \qquad (9)$$

For each Technology;

$$P_{RRU} = P_{RRU\,Sector\,1} + P_{RRU\,Sector\,2} + P_{RRU\,Sector\,3} \qquad (10)$$

As mentioned earlier, this analysis focuses on the traffic dependent components, this means that the study focuses on equations (9) and (10). The only parameter that this study analyses for comparison is the $P_{RRU}$.

#### 2.2.4.2 Validation of data using Linear Regression model

Regression models are generatedin order to determine the validate these data. As a function of traffic load given by Equations (5), the built regression models express each BTS's power usage as a function of traffic load (James D. 2016).

$$Y = a + x\beta \qquad (11)$$

Where;
Y = average power consumption (dependent variable)
a = regression coefficient (constant term)
β = weight of corresponding (coefficients)
X = independent variable (traffic load)

Assuming that the measured power P [kW] is the response Y and that the average traffic Tr [Erl] is an independent variable X. The intercept is represented by the regression line's coefficient, the constant term is represented by. Potential for a typical regression.

The developed linear regression models validate the data for accuracy using R-squared values for each model, which is also utilized to correlate the variables. The R-square range demonstrates how well a statistical model or regression line fits the actual values of the scattered real data and hence validates the data. Therefore, it is very helpful to anticipate the necessary power level to meet the continuously rising traffic demand using the developed regression models.

### 3.0 Results and Discussions
### 3.1 Traffic vs Power consumption

Measured power consumption values were measured versus their corresponding traffic load values in this section. A linear relationship between power consumption and traffic load is seen in the examples below for GSM 900, GSM 1800, and UMTS 2100. Since the measurement covered a full month, or 4 weeks, we provide some of the measured values for power consumption and traffic load in the following figures below.

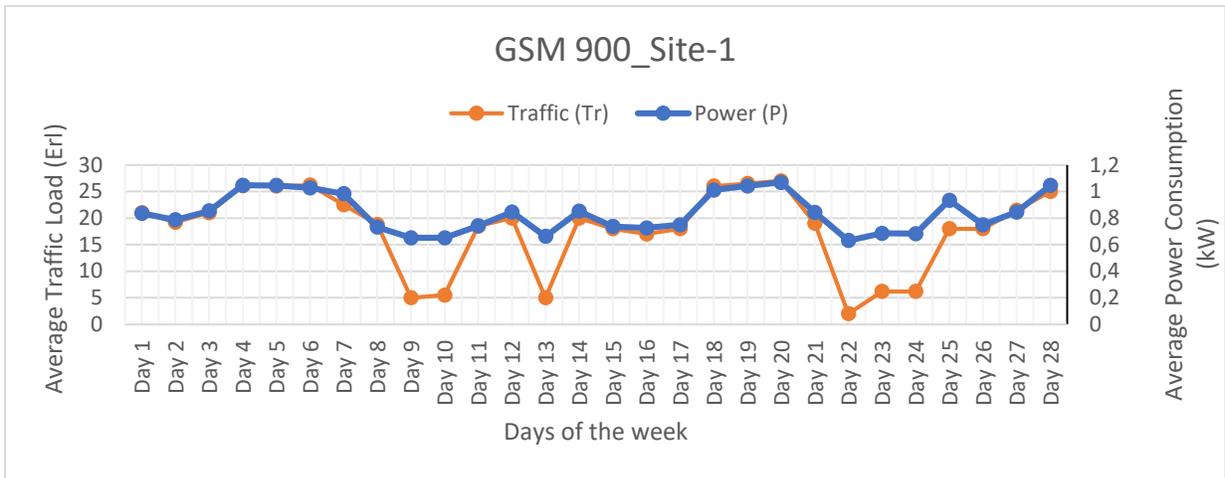

**Figure 4** GSM 900 Power consumption vs Traffic load Site-1.

The lowest power consumption is for 900 MHz is 0.62 kW and the peak value for power consumption is 1.1 kW.

From day 4 up to day 7, power consumption goes up to 1.09 kW. But from day 8 up to day 17 power consumption drops to an average of 0.65 kW. From day 18 up to day 20, power consumption rises to an average of 1.05 kW. Then begins to drop constantly until the graph is finished.

The consumption of power depends totally on the traffic load. The more call made on the network; the more power is consumed.

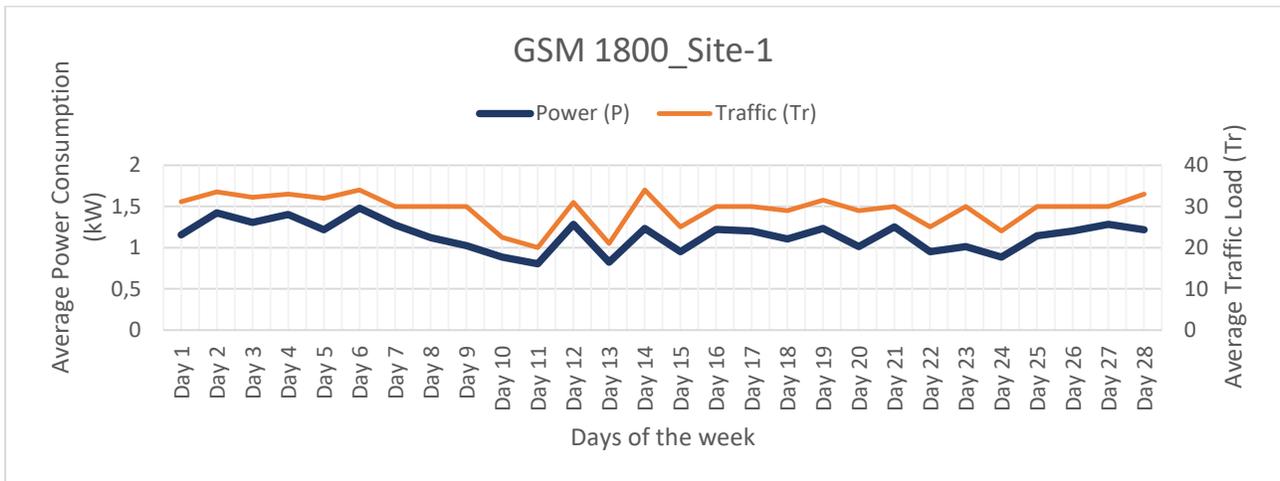

**Figure 5** GSM 1800 Power consumption vs Traffic load Site-1.

The lowest power consumption is for 1800 MHz is 0.80 kW and the peak value for power consumption is 1.5 kW.

From day 2 up to day 6, power consumption goes up to the average of 1.4 kW. But from day 7 up to day 11, then power is unstable depending on the number of calls made over that period. From day 16 power consumption stabilizes at an average of 1.2 until the end of the graph. This implies that interdependence of the traffic calls over the consumption is true.

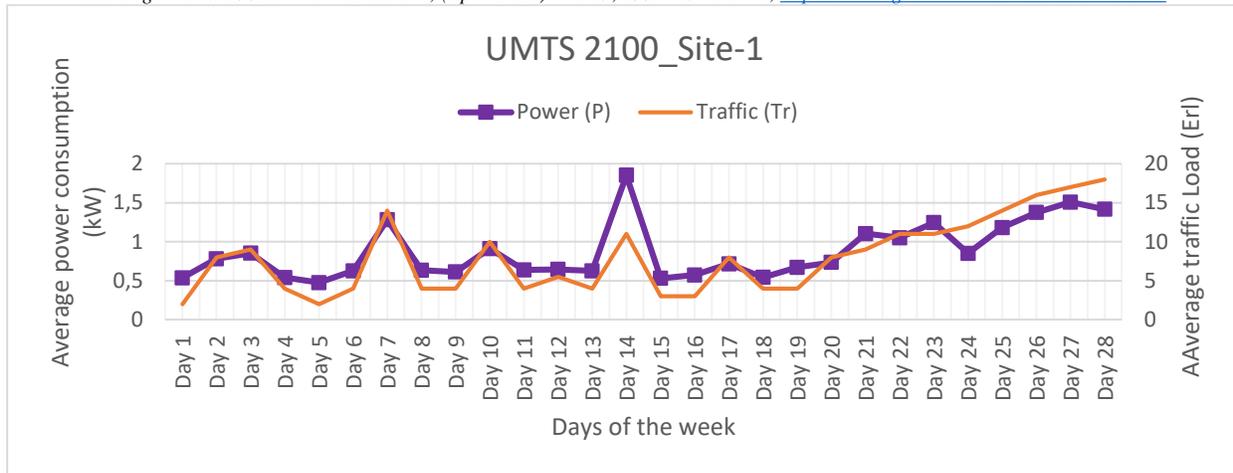

**Figure 6** UMTS 2100 Power consumption vs Traffic load Site-1.

The lowest power consumption is for 2100 MHz is 0.5 kW and the peak value for power consumption is 1.8 kW.

From day 1 up to day 6, power consumption varies at an average of 0.5 kW. But on day 7 up the consumption changes to 1.4 kW, likewise on day 14, the consumption goes up to 1.8 kW, then power is stabilized and varies according to the corresponding traffic load until the graph ends.

This fig. 6 is a typical graph where energy consumption is not stable and needs to be regressed in order to obtain a stable power consumption so as to meet the standard under which this equipment is supposed to operate. This means that this graph needs to be modeled so as to validate its consumption at site.

## 3.2. Power consumption vs traffic load regression model

To validate our data, models were developed for each analysis. As a result, the values of the models were compared and validated using the R-squared values. The obtained results are all above 0.5, this means that the model is valid. The R-squared values, indicating a similar precision sample. This model was developed in MATLAB R2021a.

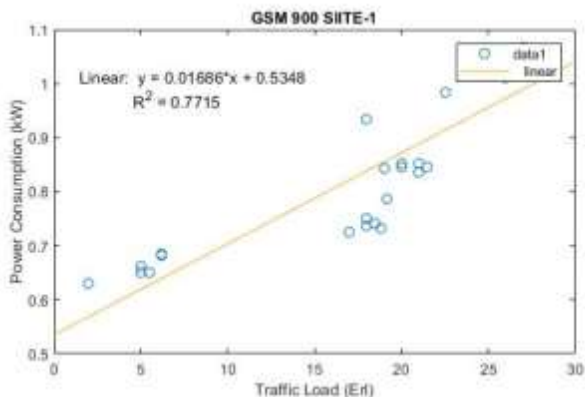

**Figure 7** Regression Model of site-1 GSM 900.

Values scattered are well arranged in this model, this means that if we use this model to implement the power consumption, we have possibility of 77% that the consumption will be linear and smooth. The line equation is given by Y = 0.01686*Tr + 0.5348.

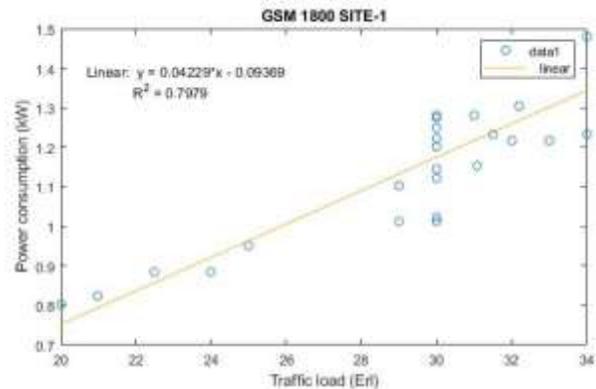

**Figure 8** Regression Model of site-1 GSM 1800.

From Fig. 8, we realize that data values are well arranged on the regression and fit this regression with an R-Square value of 0.7979. This validates the data and makes this model a perfect one for linear and stable energy consumption.

Predictions of how power would be consumed can be made, if values are substituted in the line equation, which is given by the following equation: Y = 0.0422*Tr – 0.09369. By replacing the values of Tr (Traffic load) we can be able to predict the power consumption of these base stations.

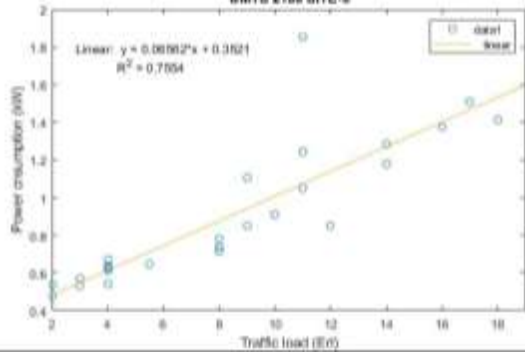

**Figure 9** Regression Model of site-1 UMTS 2100.

From Fig. 9, data values are well arranged on the regression with an R-Square value of 0.7554. This validates the data and makes this model a perfect one for linear and stable consumption.

Predictions of how power would be consumed can be made, using the line equation of the model given by; Y = 0.06562*Tr + 0.03521. By replacing the values of Tr (Traffic load) we can predict the power consumption of these base stations.

## 4.0    Discussions

For various traffic patterns, the energy efficiency of base stations installed in both urban and rural locations is examined. Equation (2.9) describes the daily average energy usage for each rural and urban base station for a period of twenty-eight days. Figures 4 – 6 display the correlation between each base station's power usage and traffic load. Regression models have been developed to validate these data by the careful monitoring of actual power consumption vs traffic demand for 28 consecutive days. Based on the model, power consumption and coverage range have been used to compare the energy utilization efficiency BTS. Figures 7 – 9 show the regression models of the BTS, having two different technologies GSM and UMTS with 3 frequency bands, 900, 1800 and 2100 MHz. The linear model does not depict the linear relationship at some point during periods of low traffic. This implies that the developed models are substantially more advantageous when there is heavy traffic than when there is light traffic.

## 5.0    Conclusions

In this paper, the objective is aimed at improving on-site mobile networks' energy utilization efficiency. Additionally, we conduct research on how the traffic depth affects the power usage of BSs. Real-time traffic and power consumption of data collected from locations with BTSs for GSM 900, GSM 1800, and UMTS access technologies have been analyzed. Results were obtained and confirm that the average power consumption of BTSs varies in accordance with the average traffic load after twenty-eight days of non-stop observations. This means that, when there is little traffic, energy use is almost the same.

The generated regression models further demonstrated that the area and average power consumption are significantly impacted by the traffic changes for rural and urban BTSs. It was found that the power consumption of the urban BTSs was roughly twice that of the rural BSs.

## Acknowledgements

The author wishes to express his sincere gratitude to the Almighty God for this journal, grate tanks to Department of Electrical, computer and Telecommunication Engineering, Kampala internal university, thanks to all the supervisors. Thanks to Airtel and ATC Uganda for allowing this research to take place on their sites.

## Declaration of conflict of interest

All authors have participated in each activity of this journal; via drafting the article and revising it critically for important intellectual content. This journal has not been submitted to, nor is under review at, another journal or other publishing venue. The authors have no affiliation with any organization with a direct or indirect financial interest in the subject matter discussed in the manuscript.